\documentclass[fleqn,10pt]{wlscirep} 
\usepackage{color,soul} 
\usepackage[]{algorithm2e} 
\usepackage{multirow} 
\usepackage{cite}
\usepackage{rotating} 
\usepackage{caption}

\title{\centering Femtosecond CDMA Using Dielectric Metasurfaces:\\ Design Procedure and Challenges}

\author[1,+]{Taha Rajabzadeh} 
\author[1,+]{Mohammad Hosein Mousavi} 
\author[2,+]{Sajjad Abdollahramezani} 
\author[3,+]{Mohammad Vahid Jamali} 
\author[1,*]{Jawad A. Salehi} 
\affil[1]{Optical Networks Research Lab (ONRL), Department of Electrical Engineering, Sharif University of Technology, Tehran, 11365-11155, Iran} 
\affil[2]{School of Electrical and Computer Engineering, Georgia Institute of Technology, Atlanta, GA 30308, USA}
\affil[3]{Electrical Engineering and Computer Science Department, University of Michigan, Ann Arbor, MI 48109, USA}
\affil[*]{Email address of the corresponding author: jasalehi@sharif.edu} 
\affil[+]{These authors contributed equally to this work}

\begin{abstract} 

Inspired by the ever-increasing demand for higher data transmission rates and the tremendous attention toward all-optical signal processing based on miniaturized nanophotonics, in this paper, for the first time, we investigate the integrable design of
coherent ultrashort light pulse code-division multiple-access (CDMA) technique, also known as femtosecond CDMA, using all-dielectric metasurfaces (MSs). In this technique, the data bits are firstly modulated using ultrashort femtosecond optical pulses generated
by mode-locked lasers, and then by employing a unique phase metamask for each data stream, in order to provide the multiple access capability, the optical signals are spectrally encoded. This procedure spreads the optical signal in the temporal domain and
generates low-intensity pseudo-noise bursts through random phase coding leading to minimized multiple access interference. This paper comprehensively presents the principles and design approach to realize fundamental components of a typical femtosecond CDMA encoder, including the grating, lens, and phase mask, by employing high-contrast CMOS-compatible MSs. By controlling the interference between the provided Mie and Fabry-Perot resonance modes, we tailor the spectral and spatial responses of the impinging light locally and independently. Accordingly, we design a MS-based grating with the highest possible refracted angle and, in the meantime, the maximized efficiency which results in a reasonable diameter for the subsequent lens. Moreover, to design our MS-based lens commensurate with the spot size and distance requirements of the pursuant phase mask, we leverage a new optimization method which splits the lens structure into central and peripheral parts, and then design the peripheral part using a collection of gratings converging the impinging at the subsequent phase mask. This work can be regarded as a pioneering attempt to make a bridge between the recent tremendous advancement in the nonophotonics technology and that of optical communication networks to extend the boundaries and facilitate their effective miniaturized deployment. 

\end{abstract}

\begin{document}

\flushbottom \maketitle

\section*{Introduction} 

As an inseparable ingredient of optics and photonics fields, light shaping techniques in which arbitrary ultrashort laser pulses are modulated to realize user-defined waveforms has gained significant interests during the past
decades. Recent advancements in the cutting-edge technologies including scanning microscopy, direct laser writing/cutting, imaging, and signal processing stem in the development of novel approaches of beam sculpting \cite{cundiff2010optical, weiner2011ultrafast}. More importantly, optical communications with distinctive capabilities such as bit error correction, electromagnetic (EM) interference immunity, high data transfer rates, and secrecy are becoming more and more attractive
\cite{rubinsztein2016roadmap}. Although a remarkable growth has been made with the multiplexing, demultiplexing, and correlation methods in the physical layer, a challenge with the developed systems still exists. Beam manipulation in the hardware level of
all-optical communication systems are majorly dependent on conventional EM components suffering from bulky geometries which hamper their merging with the ubiquitous integrated photonics. Considering the late advancement in nanotechnology fabrication,
realizing ultra-flat, highly-integrable, and wave-based photonic devices have recently garnered more attentions \cite{silva2014performing,yu2014flat}. To turn the miniaturized communication systems into the reality, evolution of novel encoding/decoding
techniques, capable of directly manipulating the information in the optical domain along with ultra-compact flat optical devices, to control the impinging wave dynamics in the subwavelength regime, are inevitable.

As a mature spread-spectrum method in the optical networks, code-division multiple-access (CDMA) which provides simultaneous, (a)synchronously multiplexed, and random channel access to a large number of users has gotten much more attentions among other
techniques \cite{salehi1989code1, salehi1989code2}. Ultrashort light pulse CDMA, also known as femtosecond CDMA, is an efficient technique where the data bits modulated using ultrashort femtosecond optical pulses, generated by mode-locked lasers, are
spectrally encoded using a unique phase mask for each data stream to provide the multiple access capability. In fact, this procedure spreads the optical signals in the temporal domain and generates low-intensity pseudo-noise bursts through random phase
coding leading to minimized multiple access interference \cite{salehi1990coherent}. The key elements in the commercialized femtosecond CDMA networks are encoders and decoders comprising of traditional components including gratings, lenses, and phase masks.
Such conventional optical devices rely on the phase accumulation effect during wave propagation in the given medium to modify the state of EM wave leading to the complex, volumetric configurations prone to the detrimental effects such as reflection losses,
alignment difficulty, and vibration sensitivity \cite{zheludev2012metamaterials, jahani2016all}. To overcome such drawbacks, a novel class of nanoscale artificial media has recently introduced yielded to the revolutionary photonic phenomena. All-dielectric
optical metasurfaces (MSs), two-dimensional versions of metamaterials, composed of (a)periodic array of subwavelength high-contrast resonators (i.e., meta-atoms), which simultaneously couple to the electric and magnetic components of the incident EM
fields, offer a promising alternative to the traditional optics \cite{arbabi2015dielectric, lin2014dielectric, staude2013tailoring, khorasaninejad2016metalenses}. Such MSs enable abrupt modifications of EM waves dynamic in terms of amplitude, phase, and
polarization offering a variety of intriguing optical effects associated with promising practical applications such as biosensing \cite{liang2017subradiant}, energy harvesting \cite{brongersma2014light}, analog computing
\cite{chizari2016analog,abdollahramezani2017dielectric}, and cloaking \cite{cheng2016all}, to name a few.

In this paper, we propose for the first time, a novel class of miniaturized, integrable, and wave-based encoder system for optical CDMA networks by incorporating multiple metadevices comprising of ultra-compact, highly-efficient, and planar all-dielectric
MSs. After a theoretical investigation of ultrashort light pulse CDMA technique, the principles and design approach to realize fundamental components of a typical encoder, including the grating, lens, and phase mask, by employing high-contrast
complementary metal oxide semiconductor (CMOS) compatible MSs will be comprehensively presented. The stunning capability of MSs in spatial, spectral, and/or temporal manipulation of the light substantiates the effectiveness of our approach in simultaneous implementation of on-chip encoders and
decoders necessary for a fully integrable nanophotonic-based CDMA network.

\section*{Theory of Femtosecond Optical CDMA Technique} 

\begin{figure}[t] 
\centering 
\includegraphics[trim=0cm 6cm 1.4cm 0cm,width=16.6cm,clip]{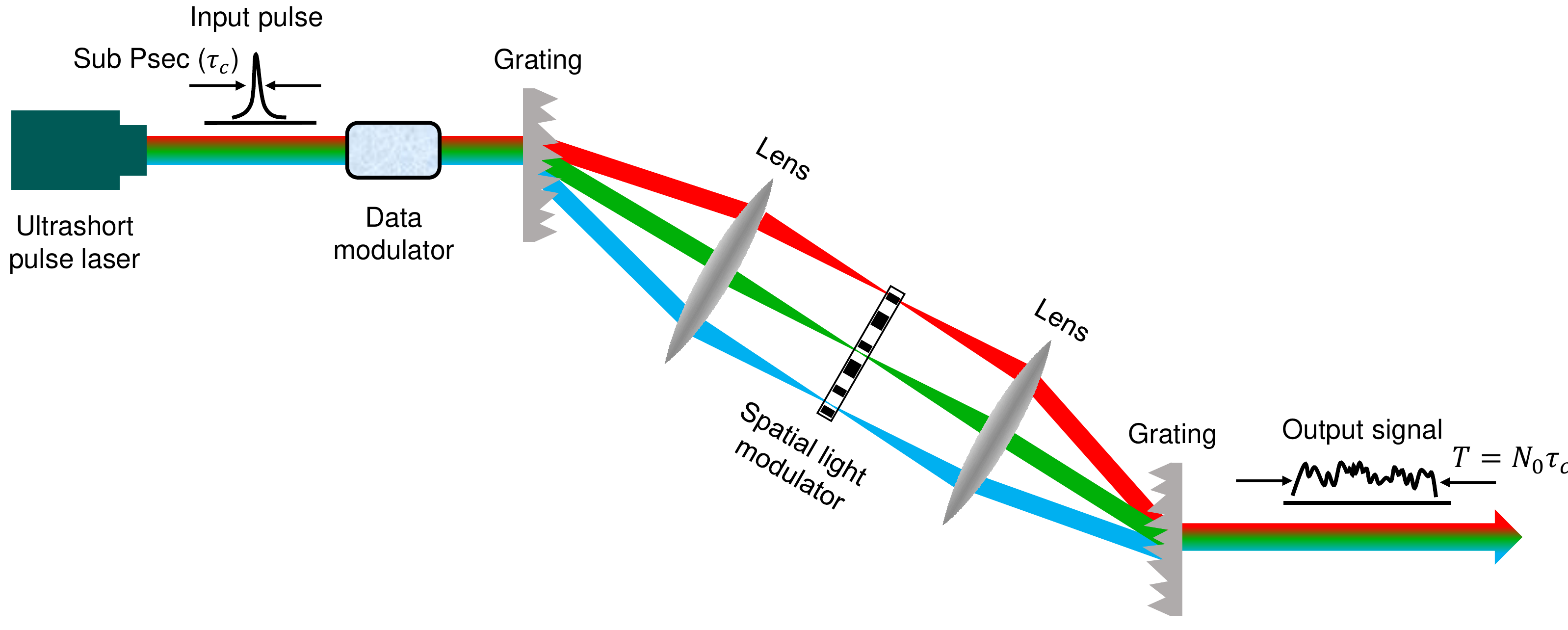} 
\caption{Schematic representation of a typical encoder of femtosecond optical CDMA network. First, the data modulator modulates the incoming ultrashort light pulses on the
 order of pico-femto seconds. Then the first Fourier transform lens focuses different spectral components of the incident light, diffracted by the first grating, into different pixels of the phase mask. At the end, the second lens and grating reassemble all
 the encoded spectral components of the spatial light modulator and construct a single phase-encoded optical beam.} 
\label{fig1} 
\end{figure}

Coherent ultrashort light pulse CDMA technique was firstly introduced by Salehi \textit{et al.} in $1990$ as an effective all-optical signal processing multiple-access mechanism in fiber-optic networks,
though using bulky optics \cite{salehi1990coherent}. In this technique, as shown in Fig.~\ref{fig1}, in order to spectrally encode a specific data stream, the coherent ultrashort pulse output of a mode-locked laser, representing one bit of information, is directed
to a grating which spatially decomposes the spectral components of the incoming ultrashort pulse with a specific resolution. Then the diffracted spectral components of the grating are all collected and focused by a lens which gives the Fourier transform of
the input light at its focal plane. At the focal plane of the lens, where different components are spatially separated, a phase mask is inserted to introduce pseudorandom phase shifts among the different spectral components. The second lens and grating
after the phase mask reassemble the encoded version of different spectral components into a single optical beam. In this case, the temporal profile of the encoded optical beam can be obtained from the Fourier transform of the transferred pattern of the
mask onto the spectrum of the incoming ultrashort light pulse; therefore, the pseudorandom phase mask temporally spreads the input ultrashort pulse into a low-intensity pseudorandom burst.
In a femtosecond CDMA network, each transmitter employs its own distinct phase mask to spectrally encode a data stream based on the above-explained procedure and then broadcasts the encoded pulses into a common optical channel, e.g., a fiber-optic channel,
that is shared between all the transmitters and receivers. The two main parts of each receiver in an ultrashort light pulse CDMA network are: the decoder which exactly has the same architecture of the corresponding transmitter encoder except that its phase
mask is the complex conjugate version of the transmitter phase mask, and the optical threshold device. Therefore, when the transmitter and receiver phase masks are a complex conjugate pair, the spectral phase shifts all components are removed and the original coherent
ultrashort pulse can properly be decoded and reconstructed. On the other hand, the undesired receivers, without knowledge of the mask architecture of the desired transmitter, cannot remove the spectral phase shifts. In other words, when the encoded signal of the
desired transmitter passes through the decoder of a typical undesired receiver, the phase masks do not match thoroughly, and hence the spectral phase shifts rearrange and do not vanish; therefore, the decoder's output signal remains a low-intensity pseudorandom
temporally-spread burst. In this case, the threshold device is set to reject any improperly-decoded low-intensity pseudorandom burst and detect the properly-decoded intense pulses of the matched transmitter-receiver pairs.

To better envision the femtosecond CDMA technique, we briefly overview the essence of the physics behind this technique. Let us consider the baseband Fourier spectrum of the starting ultrashort pulses as $A(\omega)=\frac{\sqrt{P_0}}{W}\Pi(\omega/W)$, i.e.,
a rectangular spectrum with amplitude ${\sqrt{P_0}}/{W}$ over the angular frequency band $-W/2\leq \omega\leq W/2$, where $W$ is the total bandwidth of the band-limited source and $P_0$ is the peak power of the ultrashort light pulse. Then the time-domain version
of the input signal can be expressed as $a(t)=\sqrt{P_0}~{\rm sinc}\left(\frac{W}{2}t\right)$ and consequently the corresponding instantaneous power is given by $P(t)={P_0}~{\rm sinc}^2\left(\frac{W}{2}t\right)$, where ${\rm sinc}(t)=\sin(t)/t$ is the sinc
function. We note that $P(t)$ describes an ultrashort pulse with the approximate temporal duration of $\tau_c=2\pi/W$ \cite{salehi1990coherent}.
In order to spectrally encode the starting ultrashort light pulse, $A(\omega)$ is multiplied by the Fourier response of the mask which consists of $N_0=2N+1$ distinct chips each with the bandwidth $\Omega=W/N_0$. We assume that the phase value of each
chip, which operates on a distinct spatial component of the focused spectrally decomposed light impinging on the mask, can independently be adjusted while all of the chips have the same amplitude response such that the relative amplitude of the different
spectral components remains unchanged when they pass through the mask. Considering the $n$-th chip of the mask with the unit amplitude, i.e., $A(\omega)=1$, and phase $\phi_n$, the temporal shape of the encoded field amplitude can be represented as

\begin{align}
\label{C(t)} 
C(t)={\rm sinc}\left(\frac{\Omega}{2}t\right)\frac{\sqrt{P_0}}{N_0}\sum_{n=-N}^{N}\exp\left(-jn\Omega t-j\phi_n\right)=G(t)V(t), 
\end{align} 

where $j=\sqrt{-1}$. Therefore, the encoded field amplitude after the second lens and
grating is in fact a periodic pseudo-random signal $V(t)=\frac{\sqrt{P_0}}{N_0}\sum_{n=-N}^{N}\exp\left(-jn\Omega t-j\phi_n\right)$, with period $T=2\pi/\Omega=N_0\tau_c$, modulated by a real envelope $G(t)={\rm sinc}\left(\frac{\Omega}{2}t\right)$
determining the temporal width of the encoded pulse. Obviously, for $\phi_n=0$, $\forall n$, $C(t)$ reduces to $a(t)$.
Considering binary phase masks in which the phase of each spectral chip is either $0$ or $\pi$ with the probabilities $p$ and $q=1-p$, respectively, $\phi_n$'s are independent and identically distributed (i.i.d.) random variables (RVs) with the probability
density function (PDF) $P_{\phi_n}({\phi_n})=p\delta({\phi_n})+q\delta({\phi_n}-\pi)$ where $\delta(.)$ represents Dirac's delta function. In this case, the ensemble average of $V(t)$ is given by $\langle V(t)\rangle=(p-q)V_p(t)$ where
$V_p(t)=\frac{\sqrt{P_0}}{N_0}\sum_{n=-N}^{N}\exp \left(-jn\Omega t \right)$ 
denotes the amplitude of an ideal mode-locked laser 
($\phi_n=0$ for all $n$'s) with period $T$; this means that for the random coding, i.e., $p=q=0.5$, the ensemble average of $V(t)$
is equal to zero. Moreover, based on the mathematical manipulations in Ref. \cite{salehi1990coherent}, the instantaneous power of the normalized field amplitude $C(t)$ is given by $|C(t)|^2=G^2(t)I(t)$, where $I(t)$ can be obtained as

\begin{align}
\label{I(t)} 
I(t)=\frac{{P_0}}{N_0}+\frac{{P_0}}{N^2_0}\sum_{n=-N}^{N}\sum_{m\neq n=-N}^{N}\exp\left(-j(n-m)\Omega t -j(\phi_n-\phi_m)\right). 
\end{align} 
And the ensemble average of $I(t)$ can be expressed as 
\begin{align}
\label{<I(t)>}
\langle I(t)\rangle=\frac{{P_0}}{N_0}+\frac{{P_0}}{N^2_0}(p-q)^2\sum_{n=-N}^{N}\sum_{m\neq n=-N}^{N}\exp\left(-j(n-m)\Omega t)\right)=\langle I(t){\rangle}_t+(p-q)^2[I_p(t)-\langle I(t){\rangle}_t], 
\end{align} 
where
$I_p(t)=\frac{{P_0}}{N^2_0}\sum_{n=-N}^{N}\sum_{m=-N}^{N}\exp\left(-j(n-m)\Omega t\right)$ is the intensity of an ideal mode-locked laser with period $T$ and peak power $P_0$ without a phase mask, and $\langle I(t){\rangle}_t=P_0/N_0$ is the time average
of the laser intensity over a period of $T$ seconds. While for the uncoded pulses ($p=1$) the ensemble average of $I(t)$ is equal to $I_p(t)$ which reflects an ultrashort pulse with the temporal duration $\tau_c$ and the peak power $P_0$, for the random
coding ($p=q=0.5$) the ensemble average is $\langle I(t){\rangle}=\langle I(t){\rangle}_t=P_0/N_0$. Therefore, in the case of random coding, the ideal ultrashort light pulse $I_p(t)$ disappears and the output average intensity is reduced by a factor of
$N_0$ due to the intensity spread over an $N_0$ times larger time period. Clearly, for the other values of $p$ the peak of $\langle I(t){\rangle}$ is between the peak power of the uncoded pulse ($P_0$) and that of the randomly encoded pulse ($P_0/N_0$).
Further statistical analysis in Ref. \cite{salehi1990coherent} have revealed that for the randomly encoded ultrashort light pulse with an unknown transmission time and initial phase, the real and imaginary parts of the encoded signal are statistically
independent joint Gaussian RVs. Therefore, the encoded intensity signal has a chi-square PDF as 

\begin{align}	
{\label{chi square}} 
P_I(I)=\frac{N_0}{P_0}\exp\left(-\frac{IN_0}{P_0}\right),~~~~ I\geq 0, 
\end{align} 
which is a typical PDF for all polarized
random light \cite{mandel1958fluctuations,mandel1959fluctuations,mandel1965coherence,loudon2000quantum,goodman2015statistical}. Therefore, it is claimed that the encoded femtosecond pulses appear as random light to a typical undesired receiver which has no
knowledge of the code, transmission time, and the initial phase of the desired data stream.

\section*{Results} 

\subsection*{Design Principles and Challenges}

Before explaining our proposed design scheme, we first describe the basic principles of the integrable platform of femtosecond CDMA based on dielectric MSs. Then we highlight the main
challenges playing critical roles toward the effective design of this technique that can better envision the advantages and potentials of our proposed solution in the next subsection. To begin with, we recall the schematic configuration shown in Fig. \ref{fig1}
where the collimated ultrashort beam of a mode-locked laser impinges on the grating, which decomposes the different spectral components of the incoming light to different angles. Then each spectral component is focused, with a specific spot size, deemed
$d_{ss}$, to the mask which introduces a certain phase to each component. The phase mask can be implemented by employing a two-dimensional array of predefined unit cells; the area of each unit cell determines the spot size that the precedent lens is supposed to
produce, and the distance between each pair of two consecutive unit cells defines the center-to-center distance of the adjacent focused beams, namely $d_{sd}$. The parameter $d_{sd}$ is related to the angular distance of two adjacent diffracted spectral
components $\Delta \theta$ as 

\begin{align}
\label{d_sd} 
d_{sd}=f\Delta \theta, 
\end{align} 
where $f$ is the focal length of the lens which can be determined using the distance of two consecutive spots $d_{sd}$ when $\Delta \theta$ is known for a grating.
To evaluate the required $\Delta \theta$ for a grating in a femtosecond CDMA system, we first note that based on the generalized Snell's law of refraction \cite{yu2011light,kildishev2013planar}, for any normal incident light to the grating the refracted angle $\theta_r$ is
related to the phase discontinuity function $\Phi(x)$ as $\sin(\theta_r)=\frac{\lambda}{2\pi}\frac{d\Phi(x)}{dx}$, which means that arbitrary angles for the refracted beam can be achieved if an appropriate gradient of the phase discontinuity along the
interface $\frac{d\Phi(x)}{dx}$ is satisfied. Based on the grating phase profile, we have ${d\Phi}/{dx}=2\pi/\Gamma$, where $\Gamma$ is the grating period. Therefore, we find for the refracted angle of grating that $\sin(\theta_r)=\lambda/\Gamma$ and taking
a differentiation with respect to $\theta_r$ leads to $\Delta \theta \cos(\theta_r)=\Delta \lambda/\Gamma$. Consequently, we have 

\begin{align} 
\label{del tet} 
\Delta \theta=\tan(\theta_r)\Delta \lambda/\lambda 
\end{align} 
where $\Delta \theta$ represents angular distance of two adjacent diffracted spectral
components in which $\Delta \lambda=B/N_0$ and $B$ are the bandwidth of each spectral component impinging on the phase mask and the total bandwidth of the femtosecond CDMA system, respectively. Therefore, the required
$\Delta \theta$ can be obtained considering the design parameters of the femtosecond CDMA system including $B$, {$N_0$}, and $\lambda$, and then the desired focal length of the lens can be calculated from the distance of adjacent focused beams using Eq.
\eqref{d_sd}. On the other hand, when $f$ and $\lambda$ are known, the required diameter of the lens $D$ can be obtained based on the spot size of each focused spectral component, which itself is imposed by the dimensions of each unit cell of the phase mask,
using the relation\cite{teich1991fundamentals}

\begin{align}
\label{dss} 
D=\frac{\pi}{4\lambda f}d_{ss}. 
\end{align}
The above-discussed formulas play a critical role in the integrable design of the femtosecond CDMA system and clarify the rigorous relations between the three essential parts of the encoder, i.e., the grating, lens, and mask. To better elucidate
this relationship and the challenges they impose on the MS-based design of the femtosecond CDMA system, in the following, we provide an example using the typical parameters reported in the literature. Let us, similar to Ref.
\cite{sardesai1998femtosecond}, consider ultrashort light pulses of duration $\tau_c=510~{\rm fs}$, corresponding to the source bandwidth of $W=12.32~{\rm THz}$, encoded by a $31$-elements pseudo-random binary phase mask. In the central wavelength of
$\lambda=1550~\rm{nm}$, this is equivalent to the total wavelength bandwidth of $B=98.8~{\rm nm}$, and the per spectral component bandwidth of about $\Delta \lambda\approx 3.2~{\rm nm}$. Based on Eq. \eqref{del tet}, for a specific wavelength, the larger the
refracted angle of grating $\theta_r$ we have, the more the angular distance of different spectral components $\Delta \theta$ we get. According to what reflected in the literature, realizing MS-based gratings with angles of refraction beyond
$20^{\circ}$ significantly degrades the system performance. In this case, if we limit ourselves to the refracted angle of $\theta_r=20^{\circ}$, the per spectral component bandwidth of $\Delta\lambda=3.2~{\rm nm}$ results into the angular distance of
$\Delta \theta=0.0431^{\circ}$. On the other hand, we assume that the phase mask is composed of unit cells of with periodicity $P=500~{\rm nm}$ (details on the structure of our MS-based mask can be found in the next subsection), which imposes the spot sizes on the
order of $d_{ss}\approx 500~{\rm nm}$ and the center-to-center spot distances as large as $d_{sd}\approx 500~{\rm nm}$. These values necessitate lens focal length and diameter as high as $f=0.6654~{\rm mm}$ and $D=2.6264~{\rm mm}$, respectively, which make the integrable design of femtosecond CDMA using the available MS devices questionable. Therefore, for the integrable implementation of the femtosecond CDMA technique, it is essential to design gratings addressing highest possible refracted angles, one of the goals governing the content of this paper. For example, if we somehow design gratings with refraction angles as high as $\theta_r=75^{\circ}$, the required angular distance, focal length, and lens
diameter will be $\Delta \theta=0.4414^{\circ}$, $f=64.8939~{\rm \mu m}$, and $D=256.1392~{\rm \mu m}$, respectively.

\subsection*{Proposed Design Scheme} 
\subsubsection*{Mask} 
As the heart of a typical encoder in a femtosecond CDMA system, pulse shaping mask manipulates the dynamic of Fourier-transformed spectral components of input laser beam by introducing pseudorandom spatial phase shifts. Based on the theoretical arguments in the previous section, we first generate a pseudorandom sequence with equal probability for the phase values $0$ and $\pi$, and then embed this phase sequence into the mask. Accordingly, the transmitter side broadcasts encoded pulses experienced a phase mask with specific predefined spatial pattern as its finger print. This ensures secured communication as long as a perfect matching occurs with the decoder of the receiver side containing a conjugate of the encoding mask.
Conventional shaping masks are mostly based on spatial light modulators, acousto-optic modulators, liquid crystal modulators, and deformable mirrors which are implemented 
through microlithographic patterning techniques making the ultra-compact realization of encoders impossible \cite{weiner2000femtosecond}. The solution of such fundamental challenge goes 
through all-dielectric MSs containing nanoscale high contrast constituents capable of fully molding the impinging light in a subwavelength distance \cite{qiao2017recent,Baranov:17}. Such dielectric MSs also address the fundamental drawbacks of their plasmonic counterparts in terms of high inherent Ohmic losses, 
heating, and incompatibility with CMOS fabrication processes. All-dielectric nanoresonators provide a compelling route to highly control the transmitted light with full phase-agility ($2\pi$) based on simultaneous excitation of magnetic dipole (first Mie resonance), electric dipole (second Mie resonance), and standing wave along the particle (Fabry-Perot resonance). Such ideal resonance modes empower MSs to realize a perfect transmissive Huygens’ surface, with zero total reflectance, over the telecommunication spectral range \cite{jahani2016all}. Indeed, the engineered nanoparticles in terms of geometry dimensions, shape, period, orientation, and material tailor the spatial distribution of phase discontinuities both locally and independently over a subwavelength surface \cite{kuznetsov2016optically}.  
It is worth mentioning that the unique capacity of all-dielectric MSs in concurrently tailoring the phase and polarization of the incident beam facilitates implementation of more complicated encoder systems essential for highly-secured and invulnerable communication networks, which are vital for military applications.
Moreover, the rapidly growing field of reconfigurable all-dielectric MSs have shown great promises for realization of light-assisted tunable platforms necessary for the next-generation real-time all-optical modulators excluding electro-optic effects to get higher data transfer rates and bandwidth \cite{wang2016optically, chu2016active, wuttig2017phase, makarov2017light, cheng2017chip, raeis2017metasurfaces}.

\begin{figure}[t] 
	\centering 
	\includegraphics[trim=0cm 1.5cm 8cm 0cm,width=8.3cm,clip]{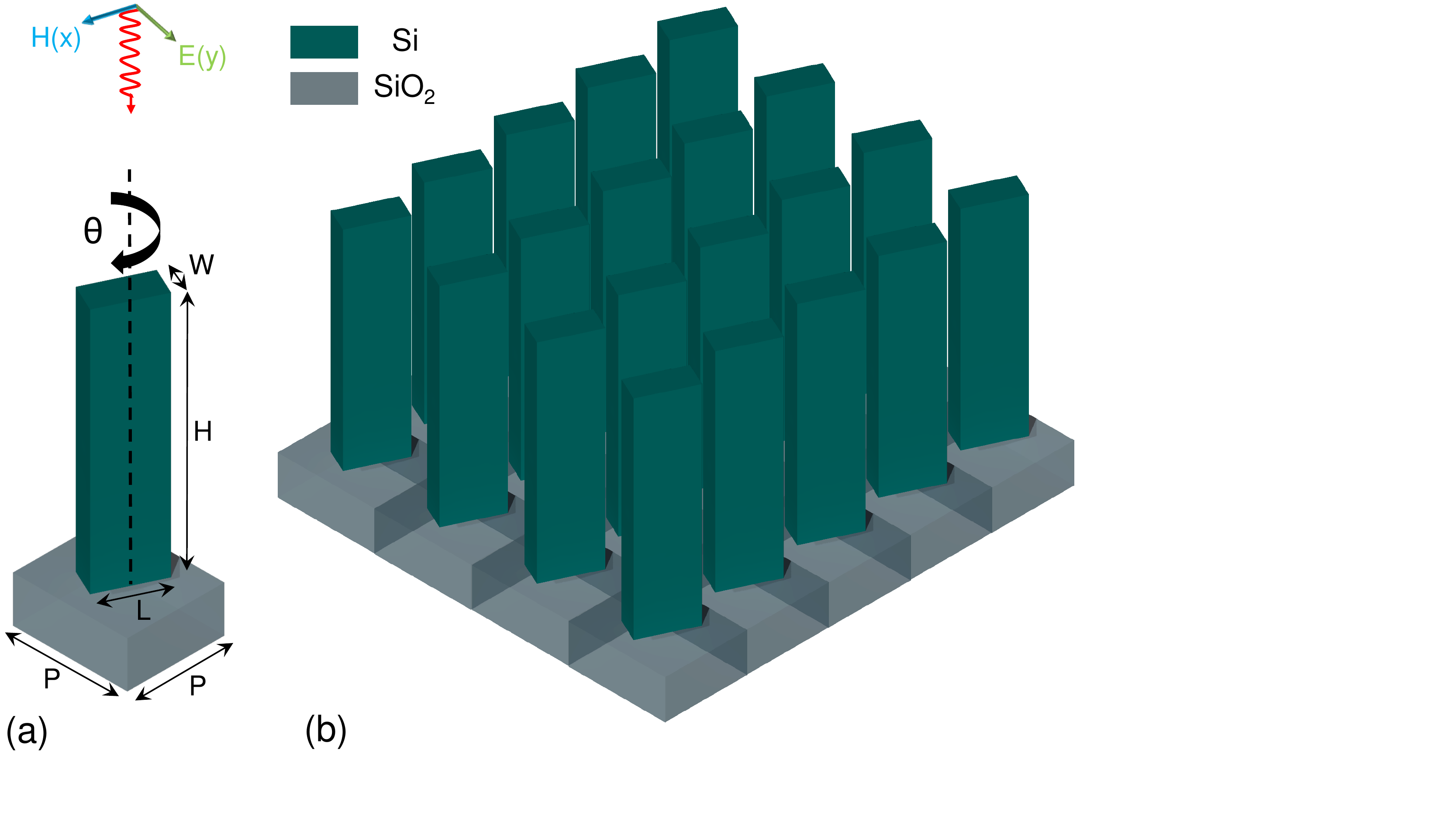} 
	\caption{ (a) Basic building block of the MS made of a Si nanofin with $H=1500~$ nm and $P=500~$ nm located on SiO$_{2}$ substrate. Other parameters are optimized based on a specific functionality of MSs in the encoder system. (b) Schematic demonstration of the proposed MS to realize main components of an optical CDMA encoder system including gratings, lenses, and mask.} 
	\label{fig2} 
\end{figure}

\subsubsection*{Grating}

\begin{table}[b]
	\begin{center} 
		\begin{tabular}{|c|c|c|c|c|c|c|c|} \cline{3-8} \multicolumn{2}{c|}{} & \multicolumn{6}{c|}{Design parameters} \\ \cline{3-8} \multicolumn{2}{c|}{} & $x$ (nm) & $y$ (nm) & $w$ (nm) & $l$ (nm) & $H$ (nm) & $\theta^{\circ}$ \\
			\hline \multirow{4}{*}{\begin{sideways} Initial~ \end{sideways}} & nanofin 1 & -850 & 0 & 160 & 380 & 1500 & 0 \\ \cline{2-8} & nanofin 2 & -270 & 0 & 180 & 330 & 1500 & 30 \\ \cline{2-8} & nanofin 3 & 0 & 0 & 90 & 280 & 1500 & 60 \\ \cline{2-8} & nanofin
			4 & 850 & 0 & 60 & 260 & 1500 & 90 \\ \hline \multirow{4}{*}{\begin{sideways} Final~ \end{sideways}} & nanofin 1 & -826 & -8 & 161 & 371 & 1500 & -0.3 \\ \cline{2-8} & nanofin 2 & -270 & 1 & 175 & 315 & 1500 & 28.2 \\ \cline{2-8} & nanofin 3 & 282 & -5 &
			116 & 281 & 1500 & 65.1 \\ \cline{2-8} & nanofin 4 & 791 & 7 & 76 & 272 & 1500 & 93 \\ \hline 
		\end{tabular} 
	\end{center} 
	\caption{Initial and optimized geometrical dimensions as well as exact position of dielectric nanoresonators in a super cell comprising of 4 unit cells necessary for realizing a grating with diffraction angle $\theta_{r}=45^{\circ}$. In the optimization process $H$ and $P$ remain constant.} 
	\label{table1} 
\end{table}

\begin{figure}[t] 
\centering 
\includegraphics[trim=0cm 8.7cm 1.4cm 0cm,width=16.6cm,clip]{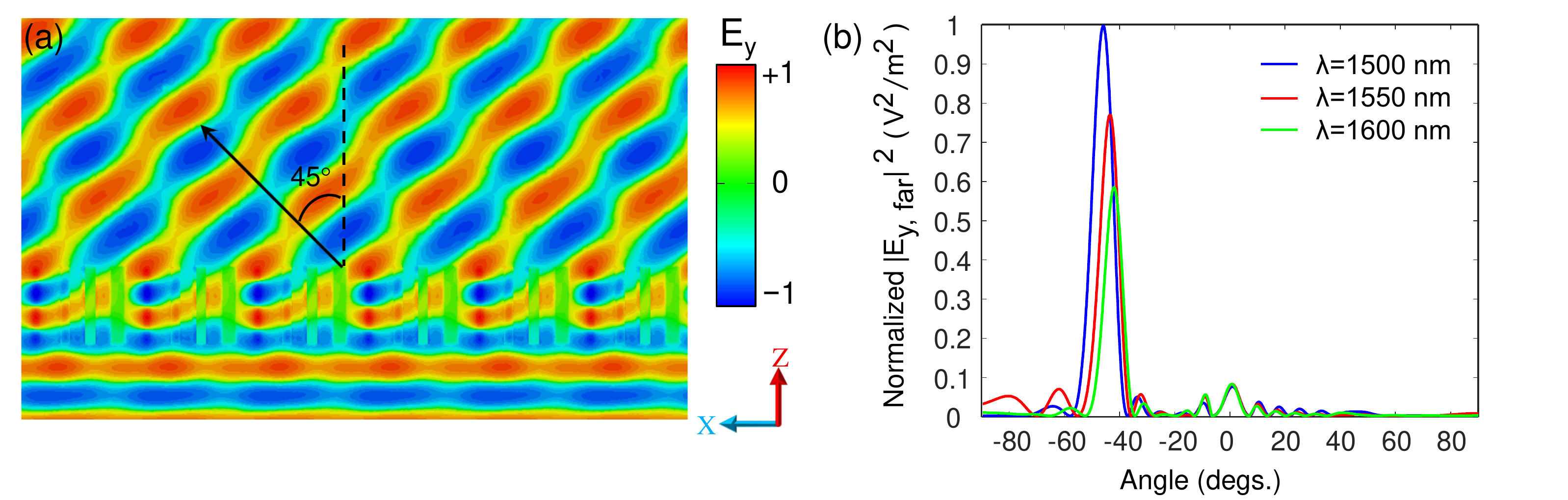} 
\caption{(a) Electrical field distribution of the incident and refracted waves at $\lambda=1550~$nm. (b) Far-field radiation pattern of the refracted beam for different input wavelengths.} 
\label{fig3} 
\end{figure}

As discussed before, the more the refracted angle the grating has, the smaller the lens we will have;
however, increasing the grating refracted angle will simultaneously decrease its efficiency. Therefore, we aim to design a grating with the highest possible refracted angle and, at the same time, with an optimum efficiency. A well-established approach
toward designing a grating is to discretize the phase profile of the grating functionality and assign a specific unit cell to each of the digitized values\cite{yang2014dielectric,decker2015high}. This can cause a significant degradation on the overall performance of gratings associated with high
refracted angles in that such a sampling method deteriorates the inherent periodicity of the sawtooth phase profile of the grating. Furthermore, employing well-known optimization methods to the above-mentioned approach is ineffective
since all unit cells are considered isolated, giving no information regarding the mutual interactions between the neighboring unit cells. To design a highly-efficient grating, a recently developed method modifies the design parameters of a building block,
namely geometrical dimensions, position, shape, and orientation angle by making the output intensity of the first diffraction order of the target refracted angle as a criteria \cite{byrnes2016designing}. Here, we utilize silicon nanofins\cite{khorasaninejad2014silicon} with a specified orientation angle as building blocks of the grating component(see Fig.~\ref{fig2}).
The design procedure of such a metasurface-based grating with the desired refracted angle and maximized efficiency is shown in Algorithm \ref{algorithm1}. As the first step, we set some nanofins with random design parameters for a specific grating period which defines the
diffracted angle according to the incident wavelength. In each step, we take one of the design parameters and gradually change it until the diffracted power of the first order is not improving anymore. Hence, in order to design a grating with this novel
algorithm, at the first, we set our target diffracted angle as $\theta_{r}=45^{\circ}$, then according to the target wavelength $\lambda=1550~{\rm nm}$ and $\sin(\theta_r)=\lambda/\Gamma$, our grating period will be $\Gamma=2.192$ $\rm{\mu m}$. Finally, by
applying Algorithm \ref{algorithm1} to four nanofins with random initial design parameters listed in Table \ref{table1}, we will reach to the optimized grating which its final parameters has been listed in Table \ref{table1} as well. The designed grating was simulated using full-wave commercial EM solver, CST Microwave Studio, and the resultant nearfiled and farfield pattens are depicted in Fig.~\ref{fig3}.

\subsubsection*{Lens}
As mentioned before, achieving practical spot size, $d_{ss}$, and distance, $d_{sd}$, on the mask necessitates diameters on the order of at least hundreds of microns for the lens, which makes the simulation process cumbersome. Design procedure of such a big lens is not feasible with conventional methods unless using supercomputers, so another approach is considered. As discussed in Ref.\cite{byrnes2016designing} to design and simulate a lens with such a size (meta-lens), we split the lens into two parts, i.e., center and periphery of
the lens. Center part of the lens is designed in a conventional method like a small lenss\cite{cheng2014wave,estakhri2014manipulating,pors2013broadband} but design of the periphery is different which will be discussed later in this subsection.

A flat lens which focuses light in the $z$-direction with focal length $f$ should be designed such that the transmit phase adopt the following hyperboloidal profile, in which $x$ and $y$ are the cartesian coordinates and $\lambda$ is our design wavelength\cite{aieta2012aberration}
\begin{align}\label{d_ss}
\phi(x,y)=\frac{2\pi}{\lambda}(\sqrt{x^{2}+y^{2}+f^{2}}-f).
\end{align}

In conventional methods, one should design a number of unit cells to span the phase shift of $0$ to $2\pi$. Then by moving in constant steps from center of the lens, and taking enough samples, the phase profile as well as place of the best-matching designed unit cells are achieved. The
unit cell in this design is a silicon nanofin with a square cross-section that by changing its length different phase shifts are achieved; hence, with this method we create our unit cell collection which consists of position of unit cells as well as their
specification for producing the phase shift profile.

\begin{figure}
	\centering 
	\includegraphics[trim=0cm 1cm 0cm 0cm,width=16.6cm,clip]{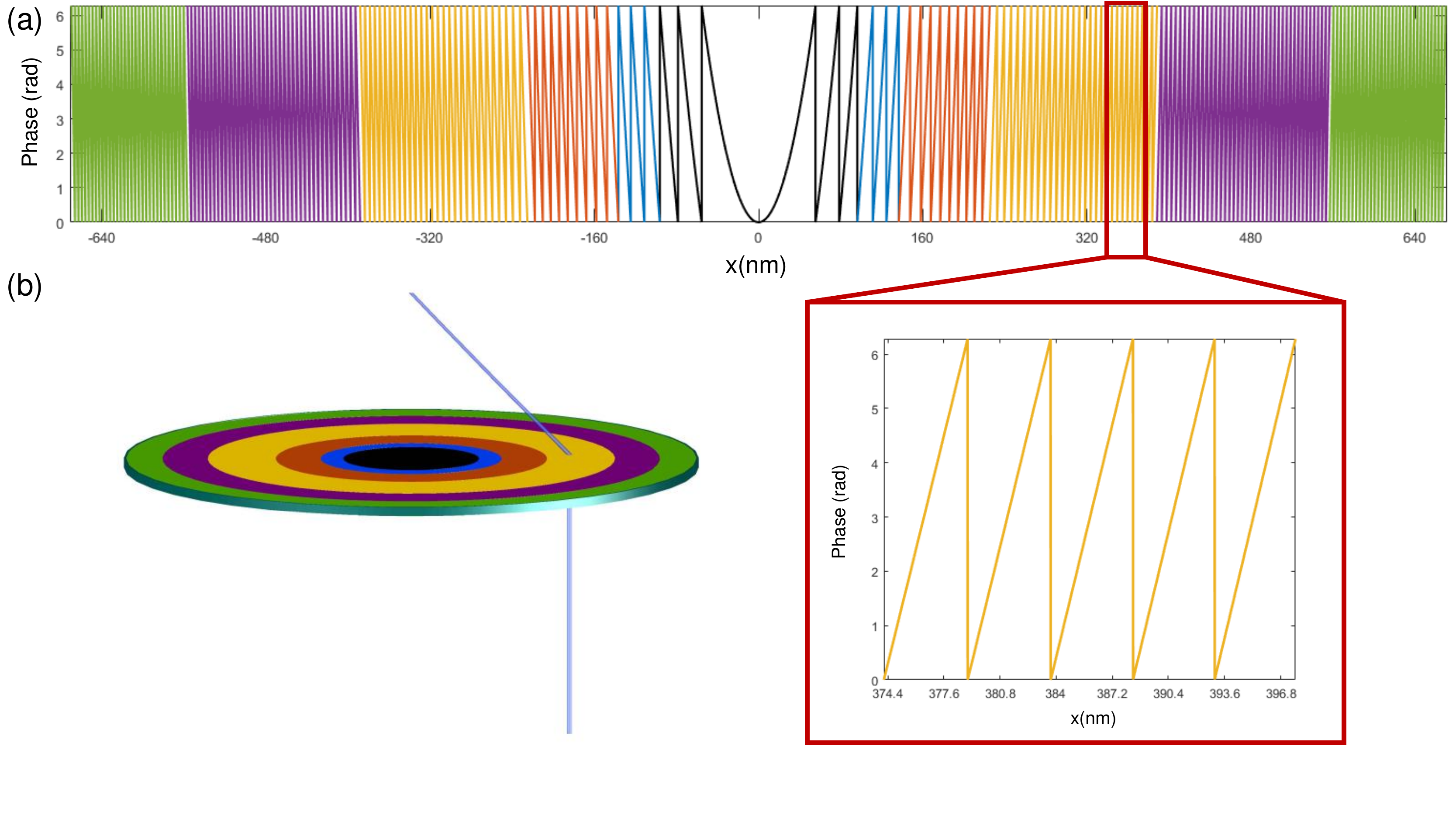} 
	\caption{Lens is divided in different parts to create phase profile of a lens. Black part is lens center and colored ones are periphery. Phase profile of each parts is shown with the same color. as demonstrated by  magnified part every colored part can be approximated to saw tooth phase profile of gratings. Each 0 to $2\pi$ phase change is a grating with certain period picked from grating archive created by method described in Algorithm \ref{algorithm2}. each colored part represents gratings from the same color section of grating archive in Fig.~\ref{fig5}.} 
	\label{fig4} 
\end{figure}

\begin{figure}[t] 
	\centering 
	\includegraphics[trim=0cm .5cm 8cm 0cm,width=8.3cm,clip]{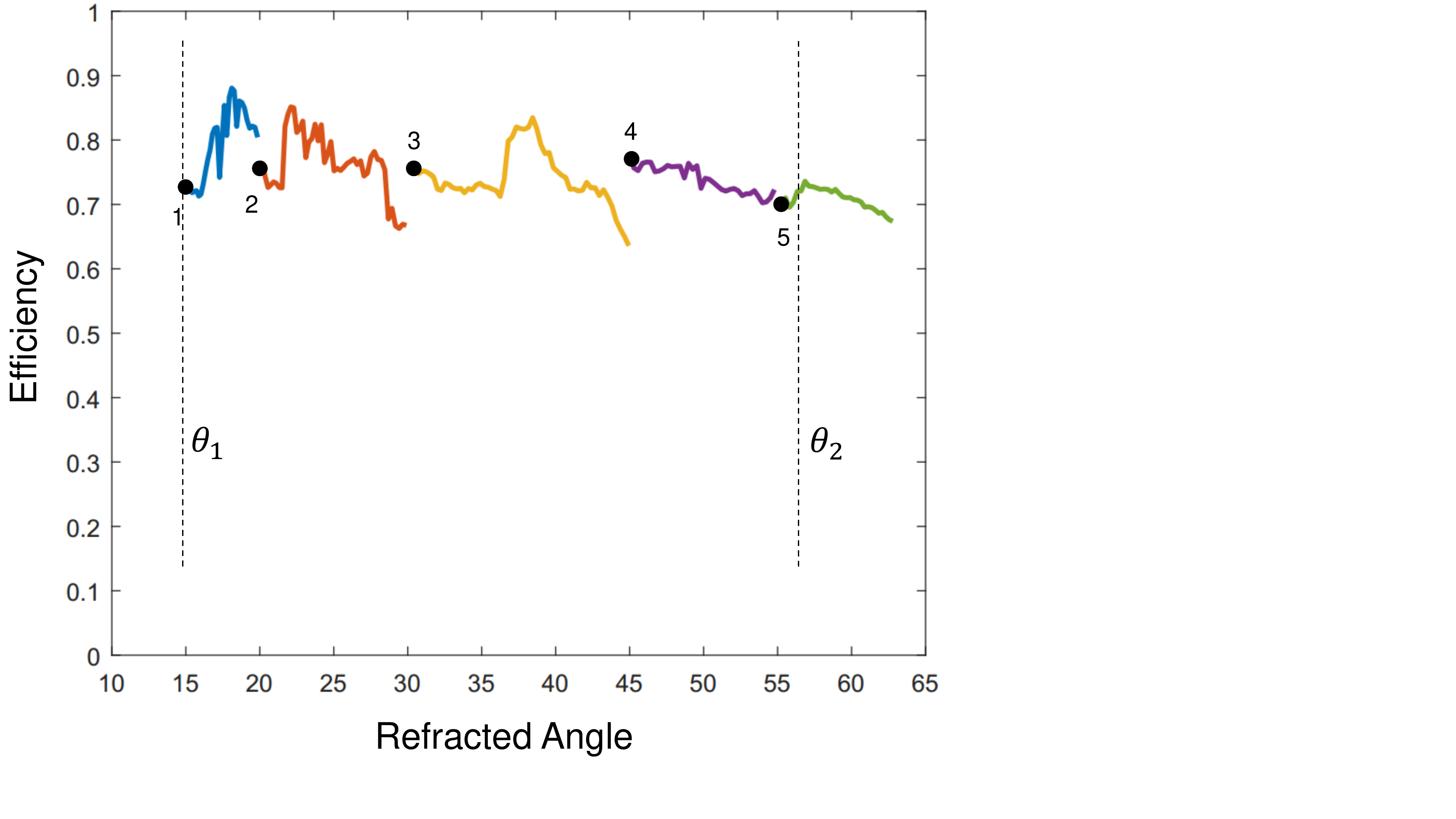} 
	\caption{Grating archive. Points 1 to 5 are the corresponding diffraction angles of initial gratings used in grating collection in Algorithm \ref{algorithm2} which are optimized separately with Algorithm \ref{algorithm1}. Starting from these gratings and with changing grating period (changing diffraction angle) gradually using the initial properties of previous step new gratings are created in order to complete the archive for every diffraction angle from $15^{\circ}$ to $63^{\circ}$. The corresponding efficiencies of every grating is demonstrated, and gratings from angle $15^{\circ}$ to $57^{\circ}$ are used in creation of meta lens.} 
	\label{fig5} 
\end{figure}

However, by getting far away from the center of lens at some point our sampling will result poorly because of the rapid changes in the phase profile which requires modifications on the design approach. Away from the lens center, hyperboloid phase profile of a lens can be approximated with a linear sawtooth phase (see Fig. ~\ref{fig4}) so that these parts can safely be assumed as gratings with slowly-changing periods and refraction angles. In other words, we approximate the periphery
of the lens with gratings which bend the incident collimated light so that the refracted light of these gratings converges at the focal length. Therefore, instead of designing every single unit cell to make the phase profile, which would be inaccurate and
impractical because of high gradient in the profile, we designed gratings with various refraction angles starting from $\theta_{1}=15^{\circ}$ (the angle at which the lens starts to be approximated via bunch of gratings) to $\theta_{2}=57^{\circ}$ (the
angle at which the lens produces the desired numerical aperture which corresponds to the desired spot size) in the way described in the previous subsection. Then for every semi-sawtooth part in the lens phase profile, we placed a grating with best matching
period in the final design of the periphery. In order to archive the gratings, we first select and optimize five gratings with the refracted angles shown in Fig.~\ref{fig5}. The optimization process is according to that of described in Algorithm \ref{algorithm1}. Then starting
with these gratings and applying the grating collection design method described in Algorithm \ref{algorithm2} the archive is complete, which is shown in Fig.~\ref{fig5}. Based on Algorithm \ref{algorithm2} to design a grating with the refracted angle $\theta+\delta\theta$ we start the optimization with the initial
values from the properties of the previously-optimized grating with the refracted angle $\theta$. Because there is only a slight difference in the period of these two gratings this initialization suggests a considerably higher starting efficiency than a
random one which significantly reduces the optimization time. Now by putting center and periphery design together our lens is completed.

The lens parameters are derived based on the specifications of the femtosecond CDMA system. In particular, similar to the previous example, we consider a femtosecond CDMA system with the total bandwidth of $100$ nm around the central wavelength of $1550$
nm comprised of a $31$-elements mask. Eqs. \eqref{d_sd}, \eqref{del tet}, and \eqref{dss} imply that a lens with $f=387.5~{\rm \mu m}$ and $R={D}/{2}=588.26~{\rm \mu m}$ is required. The farthest grating used in the lens design should have the refracted
angle of $\theta_{2}=\tan^{-1}({D}/{2f})$.

As we discussed before, the mask plays a critical role in defining the spot size of the lens meaning that our lens should produce the spot sizes equal to the size of mask unit cells, i.e., $650$ nm in our case; hence, we need to measure the spot sizes
produced by our lens. It is known that our lens focuses the collimated light into a small region in the focal plane, i.e., the focus region, and due to the reciprocity principle if we put a point source in a focus region the light which comes out from the
lens should be collimated. While we move our point source in a focus region, our far-field power will not change, but as soon as we come out from the focus region the far-field power will change. Accordingly, the distance which far-field power will change specifies the spot size of the lens. By using this method we calculate the spot size of lens about $650$ nm which is an appropriate lens for the femtosecond CDMA system.

\section*{Discussions}
In conclusion, as an attempt to incorporate the recent invaluable progress in nanophotonics technology into the advanced optical communication systems and highlight their effective miniaturized deployment, we investigated the integrable design of femtosecond CDMA technique, as a mature all-optical multiple-access mechanism in multiuser networks, using high-contrast CMOS-compatible all-dielectric metasurfaces. In particular, we rigorously developed the fundamental principles and approaches toward the realization of a typical femtosecond CDMA encoder
and explored the design of its pivotal building blocks including gratings, lenses, and phase masks. We observed that an efficacious implementation of the femtosecond CDMA system requires joint design of these essential parts meaning that each block should be designed based on the characteristics of the other blocks as well as the femtosecond CDMA system's specifications. For example, a smaller size for the lens necessitates larger angles of refraction for the precedent grating which itself decreases the grating efficiency. Therefore, we firstly designed a MS-based grating with the highest possible refracted angle and, in the meantime, the maximized efficiency which leads to a reasonable diameter for the subsequent lens. Moreover, to design our MS-based lens which fulfills the spot size and distance requirements of the following phase mask, based on an optimization procedure separating the structure of the lens into central and peripheral distinct sections, we built the peripheral part using a collection of gratings that their diffracted light converges at the focal plane of the lens where the phase mask is assumed to be placed.

This work can be considered as a pioneering attempt in extending the boundaries to bridge the recent tremendous advancement in the nonophotonics technology, specifically all-dielectric metasurfaces, to that of optical communication networks, which opens the window towards lots of interesting research activities that couple different problems in joint-optimization of optical communication networks and integrated photonics that aim to expedite the implementation of miniaturized all-optical signal processing and communication systems. Designing the decoder part of MS-based femtosecond CDMA system and then developing a rich simulation and experimental set-up to evaluate the overall performance of the underlying multiuser network can be regarded as a mature and potential future work. Moreover, joint design of different essential parts as a single compact module that performs all the functionalities together is unquestionably the next step in the integrable realization of a femtosecond CDMA system.

%\subsection*{Subsection} % %Example text under a subsection. Bulleted lists may be used where appropriate, e.g. % %\begin{itemize} %\item First item %\item Second item %\end{itemize} % %\subsubsection*{Third-level section} % %Topical subheadings are
%allowed. % %\section*{Discussion} % %The Discussion should be succinct and must not contain subheadings.
\section*{Methods}
The effective implementation of the femtosecond CDMA encoder requires joint design of the different pivotal parts of the system meaning that each block should be designed based on the characteristics of the other blocks as well as the femtosecond CDMA system's specifications. For example, a smaller size for the lens necessitates larger angles of refraction for the precedent grating which itself decreases the grating efficiency. Therefore, an optimization process is required to design the MS-based grating with the highest possible refracted angle which in the meantime achieves the maximum efficiency, leading to a reasonable diameter for the subsequent lens. 
The design procedure of such a MS-based grating which achieves the desired refracted angle with the maximized efficiency is shown in Algorithm \ref{algorithm1}. At the first, we set some nanofins with random design parameters for specific grating period which specifies the
diffracted angle according to the incident wavelength. In each step, we took one of the design parameters and gradually changed it such that the diffracted power of the first order was not improving anymore. 

Moreover, to design our MS-based lens commensurate with the spot size and distance requirements of the phase mask, we split the lens into central and peripheral parts and designed the central part using conventional methods like a small lens. On the other hand, for the design of the peripheral part, we used an optimization procedure which constructs the lens using a collection of gratings that bend the incoming collimated light in a way that the diffracted light of the gratings converge at the focal plane of the lens where the phase mask is placed. The design procedure of such a lens with the desired focal length and numerical aperture is summarized in Algorithm \ref{algorithm2}. 

\begin{algorithm} 
	%\KwData{this text} 
	\KwResult{Optimized grating with the desired refracted angle} Initializing design parameters including geometrical dimensions and orientation angle\; 
	\For{$\forall\alpha\in$\{design parameters\}}{ \While{efficiency is
			improving}{\eIf{$\alpha$ type is length}{ change it by 1 nm\; }{ change it by $0.3^{\circ}$\; } efficiency $\leftarrow$ diffracted power of the first order } } 
	\caption{Design procedure of a MS-based grating with the desired refracted angle and
		maximized efficiency.} 
	\label{algorithm1}
\end{algorithm}

\begin{algorithm}[t] 
	%\KwData{this text} 
	\KwResult{Lens with a desired focal length as well as numerical aperture} \textbf{Unit cell Collection:}\hfill{/*~\textbf{l} is size of  the unit cell cross section~*/} \\
	$l \leftarrow l_{0}$\;
	$\varphi(l)$ $\leftarrow$ calculate the
	shift phase for unit cell with cross section size of $l$\; add ($l,\varphi(l)$) to unit cell collections \; \While{unit cell collections cover $2\pi$ phase shift}{ $\varphi(l+\delta l)$ $\leftarrow$ calculate the shift phase for the unit cell with length of $l+\delta
		l$\; $l \leftarrow l + \delta l$\; add ($l,\varphi(l)$) to the unit cell collections \; } \textbf{Grating Collections:}\hfill{/*~$\theta_{1}$ is the first angle~*/} 
	\\$\theta \leftarrow \theta_{1}$\;
	g($\theta$) $\leftarrow$ design grating for $\theta$ usingAlgorithm 1\; add g($\theta$) to grating collection\; \While{$\theta<\theta_{2}$}{ g($\theta + \delta\theta$) $\leftarrow$ design grating for $\theta + \delta\theta$ using Algorithm 1 with initial g($\theta$) instead of random parameters\; $\theta
		\leftarrow \theta + \delta\theta$\; add g($\theta$) to grating collection\; } \textbf{Lens Center Profile:}\hfill{/*~$r_{0}$ is the border of center and periphery and $\Lambda_{0}$ is the step size ~*/}
	\\ 
	$n,m = 0$\;  
	\While{$\sqrt{n^{2}+m^{2}}\Lambda_{0}<r_{0}$}{ $U \leftarrow$ get unit cell from the unit cell collection with nearest phase to
		$\phi(n\Lambda_{0},m\Lambda_{0})$\; place $U$ in postion ($n\Lambda_{0},m\Lambda_{0}$) of Lens Center Profile\; increase $m$ and $n$\; } \textbf{Lens Periphery Profile:}\hfill{/*~$R$ is the lens redius~*/}
	\\
	\For{$r_{0}<r<R$}{ find $\delta r$ in a way that phase profile, $\phi$, from $r$ to
		$r+\delta r$ varies from 0 to $2\pi$\; $G \leftarrow$ get grating from grating collection with nearest grating period to $\delta r$\; place $G$ in a ring with radius from $r$ to $r+\delta r$ of Lens Periphery Profile\; }
	Final Lens Design $\leftarrow$ Lens Periphery Profile $+$ Lens Center Profile\;	
	\caption{Design procedure of a MS-based lens with a desired focal length and numerical aperture.} 
	\label{algorithm2}
\end{algorithm}

%\bibliography{sample}

%\noindent LaTeX formats citations and references automatically using the bibliography records in your .bib file, which you can edit via the project menu. Use the cite command for an inline citation, e.g. \cite{Figueredo:2009dg}.
%\section*{Acknowledgements (not compulsory)} % %Acknowledgements should be brief, and should not include thanks to anonymous referees and editors, or effusive comments. Grant or contribution numbers may be acknowledged.
\section*{Author contributions statement}
T.R. and M.H.M. carried out the research, performed simulations, and wrote Algorithms 1 and 2. S.A. and M.V.J. conducted the research and wrote the manuscript.  J.A.S. conceived the idea, proposed the research, and supervised the project. All authors discussed the results, analyzed the data, and reviewed the manuscript.
%Must include all authors, identified by initials, for example: A.A. conceived the experiment(s), A.A. and B.A. conducted the experiment(s), C.A. and D.A. analysed the results. All authors reviewed the manuscript.
%\section*{Additional information} % %To include, in this order: \textbf{Accession codes} (where applicable); \textbf{Competing financial interests} (mandatory statement). % %The corresponding author is responsible for submitting a
%\href{http://www.nature.com/srep/policies/index.html#competing}{competing financial interests statement} on behalf of all authors of the paper. This statement must be included in the submitted article file. % %\begin{table}[ht] %\centering
%\begin{tabular}{|l|l|l|} %\hline %Condition & n & p \\ %\hline %A & 5 & 0.1 \\ %\hline %B & 10 & 0.01 \\ %\hline %\end{tabular} %\caption{\label{tab:example}Legend (350 words max). Example legend text.} %\end{table} % %Figures and tables can be referenced
%in LaTeX using the ref command, e.g. Figure \ref{fig:stream} and Table \ref{tab:example}.

\begin{figure}[b] 
	\centering 
	\includegraphics[trim=0cm 9cm 10.7cm 0cm,width=16.6cm,clip]{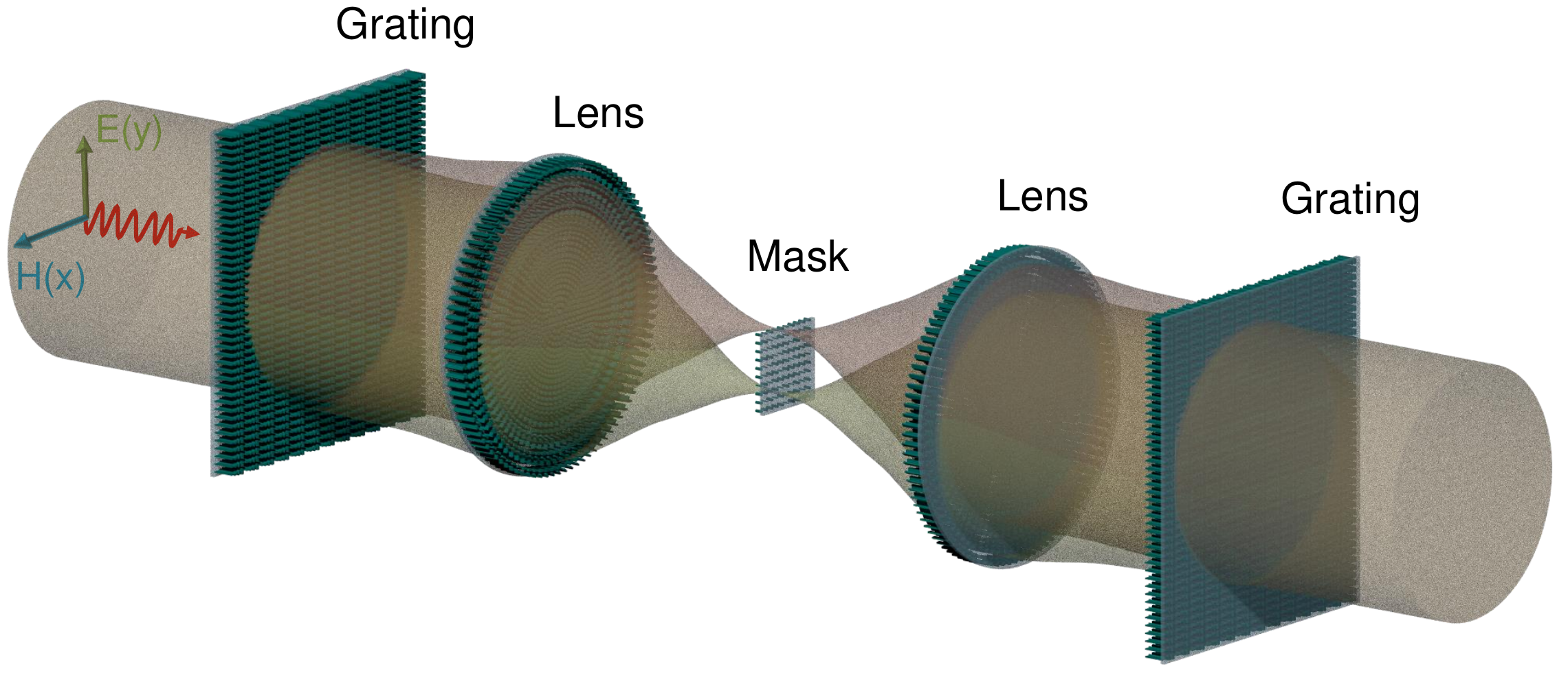} 
	\caption{Table of Content} 
	\label{fig6} 
\end{figure}

\end{document}